\documentclass[twocolumn,trackchanges]{aastex701}

\usepackage{microtype}

\begin{document}

\title{Direct observation of the X-ray counterpart of the H$\alpha$ filaments and of the sloshing spiral in the Perseus galaxy cluster}

\author[orcid=0000-0003-2251-6297]{Adrien Picquenot}
\affiliation{Department of Astronomy, Louisiana State University, LA, USA}
\email[show]{apicquenot@lsu.edu}  

\author[orcid=0000-0002-6606-2816]{Fabio Acero}
\affiliation{Université Paris-Saclay, Université Paris Cité, CEA, CNRS, AIM, 91191, Gif-sur-Yvette, France}
\affiliation{FSLAC IRL 2009, CNRS/IAC, La Laguna, Tenerife, Spain}
\email{}

\author[orcid=0000-0001-6638-4324]{Valeria Olivares}
\affiliation{Departamento de F\'isica, Universidad de Santiago de Chile, Av. Victor Jara 3659, Santiago 9170124, Chile}
\affiliation{Center for Interdisciplinary Research in Astrophysics and Space Exploration (CIRAS), Universidad de Santiago de Chile, Santiago 9170124, Chile}
\email{}

\author[orcid=0000-0002-6548-5622]{Michela Negro}
\affiliation{Department of Astronomy, Louisiana State University, LA, USA}
\email{}

\author{Gabriel W. Pratt}
\affiliation{Université Paris-Saclay, Université Paris Cité, CEA, CNRS, AIM, 91191, Gif-sur-Yvette, France}
\email{}

\begin{abstract}

Deep \textit{Chandra} observations of the Perseus galaxy cluster have allowed for the discovery of X-ray counterparts to the  H$\alpha$ filamentary structures and of a sloshing spiral. However, both components are extremely faint, and their study is largely hindered by the volume-filling hot intracluster medium (ICM). Using the Poisson General Morphological Component Analysis (pGMCA) algorithm, a blind source separation method adapted for Poissonian statistics, we were able to extract detailed, clean morphological maps of these components. We then introduced a template fitting method to investigate their spectral characteristics. We report the first direct observation of $\sim 1.35$ keV low-energy emission from the sloshing spiral, and produce the most detailed and unpolluted map of the X-rays filaments thus far obtained.

\end{abstract}

%\keywords{}

\section{Introduction}

The Perseus cluster is the X-ray brightest nearby cluster of galaxies ($z=0.018$, or about $70$ Mpc), and benefits from deep \textit{Chandra} observations \citep[][]{Fabian00,Fabian_2006,Sanders07}. Its X-ray emission is primarily due to thermal bremsstrahlung and line radiation from the hot intracluser medium (ICM), to which the  NGC 1275 galaxy contributes in the core. The deep X-ray observations have allowed for the study of the main morphological features in the core regions: the ghost bubbles (X-ray cavities) inflated by radio jets from the central galaxy \citep[][]{Boehringer93}, the sloshing spiral \citep[][]{Fabian00}, and the filamentary structures associated with colder gas that are detected in H$\alpha$ \citep[][]{Fabian03-filaments}.

The study of these distinct morphological features is key to our understanding of the formation and evolution of a cluster. Jet-inflated bubbles are clear evidence of active galactic nuclei (AGN) feedback \citep[e.g.][]{McNamara_2000,Fabian12}. Extended kiloparsec-scale filaments surround the central Brightest Cluster Galaxy (BCG), which are observed at different temperatures, from cold molecular gas, warm, and hot gas \citep[][]{salome08,tremblay18,Hamer_2019,Ubertosi_2023,Olivares_2022,olivares23,Russell19}. These filaments are believed to be condense from the hot ICM and form due to the thermal instabilities triggered by the AGN feedback. Additionally, since galaxy clusters grow hierarchically through accretion and mergers, this process leave distinct imprints in the X-ray emission, such as shock fronts, cold fronts, and signs of gas turbulence. 

However, despite the depth of the X-ray observations, the sloshing spiral and the filamentary structures in Perseus are extremely faint features whose emission is overwhelmed by that from the main ICM component. As a result, it is difficult directly to probe their thermodynamic properties. 
%compared to the appear extremely faint in X-rays, and the main features from the hot ICM have until recently hindered a direct probe of these weaker emissions. 
In \cite{Olivares_2025}, a blind source separation method allowed for the extraction of X-ray emitting filamentary structures from the \textit{Chandra} observations of seven clusters displaying prominent H$\alpha$ filaments. The surface brightness of the X-ray and H$\alpha$ filaments showed a tight correlation, indicative of a strong connection between hot and warm gas phases. This discovery provided a strong argument in favor of filamentary gas being condensed through the ICM via Chaotic Cold Accretion (CCA) \citep[e.g.][]{Sharma_2012,Wang_2021,Gaspari_2016}.

In this paper, we will use the same blind source separation method to extract clean and detailed images of the X-ray filaments and of the sloshing spiral in Perseus. We will also introduce a template fitting method inspired by gamma-ray analysis that allows us to determine the thermodynamic and chemical properties of the sloshing spiral
%the detect a low temperature $1.3$ keV emission from the sloshing spiral. 
The paper is structured as follows: in Sect. \ref{sect:analysis}, we present the analysis methods and our results, and in Sect. \ref{sect:discussion} we discuss the physical interpretation of the results.

\section{Analysis and results}
\label{sect:analysis}

\subsection{Observations}

We used the \textit{Chandra} ACIS-S observations from 2002 and 2004 (obsID 3209, 4289, 4946-4953, 6139, 6145, 6146, total exposure of $969$ ks) reprocessed with {\it CIAO 4.16.0} \citep[][]{CIAO}, and stacked the observations in a $(x,y,E)$ data cube of $288$ arcsec spatial side ($\sim~110$kpc). The brightest pixels in the central galaxy NGC 1275 were removed using a simple wavelet-based inpainting method.

For our comparison with H$\alpha$ emitting filaments, we used the \textit{WIYN 3.5-m} telescope observations first presented in \cite{Conselice_2001}.

\subsection{Spatial analysis}

To retrieve accurate maps of the H$\alpha$ filaments counterpart and of the sloshing front in X-rays, we used a blind source separation method based on the General Morphological Components Analysis \citep[GMCA, see][]{bobin15}, first introduced for X-ray observations by \cite{picquenot:hal-02160434}. An updated version, the pGMCA, was developed to take into account Poisson statistics in \cite{9215040}. The pGMCA algorithm can disentangle spectrally and spatially mixed components from an X-ray data cube of the form $(x,y,E)$ with no prior physical or instrumental information by focusing on the morphological and/or spectral diversity of the components to be separated. The pGMCA has proven capable of extracting extremely faint components from X-ray data cubes \citep[see for example][]{picquenot2024sync}.

% The pGMCA method has previously been applied to a catalog of galaxy clusters including Perseus in 
\cite{Olivares_2025} previously applied the pGMCA method to a sample of galaxy clusters, including Perseus, in order to extract X-ray counterparts to the H$\alpha$ filaments. Here, we present a better-resolved morphological map of the X-ray filamentary structures; instead of a surface brightness comparison, we use the native \textit{Chandra} resolution of this new map to study more precisely the co-spatiality of X-ray and H$\alpha$ filaments. We also present the first detailed, unpolluted map of the inner cold front and spiral structure in Perseus.

%We compare the widths of the X-ray filaments to those seen in H$\alpha$ along ten chosen profiles in Fig. \ref{fig:filaments-boxes-profiles}.

In order to increase the statistics in each pixel and obtain more accurate results with pGMCA, we rebinned the \textit{Chandra} data cube in two different ways to focus on different features. For the first pGMCA run, we rebinned spatially to a $4$" bin size, but kept the native $14.6$" eV energy bin size. On a second run, we used data rebinned spectrally to $116.8$ eV energy bins but keeping the native $0.5$" spatial bin size.

\begin{figure}[ht!]
\centering
\includegraphics[width=9cm]
{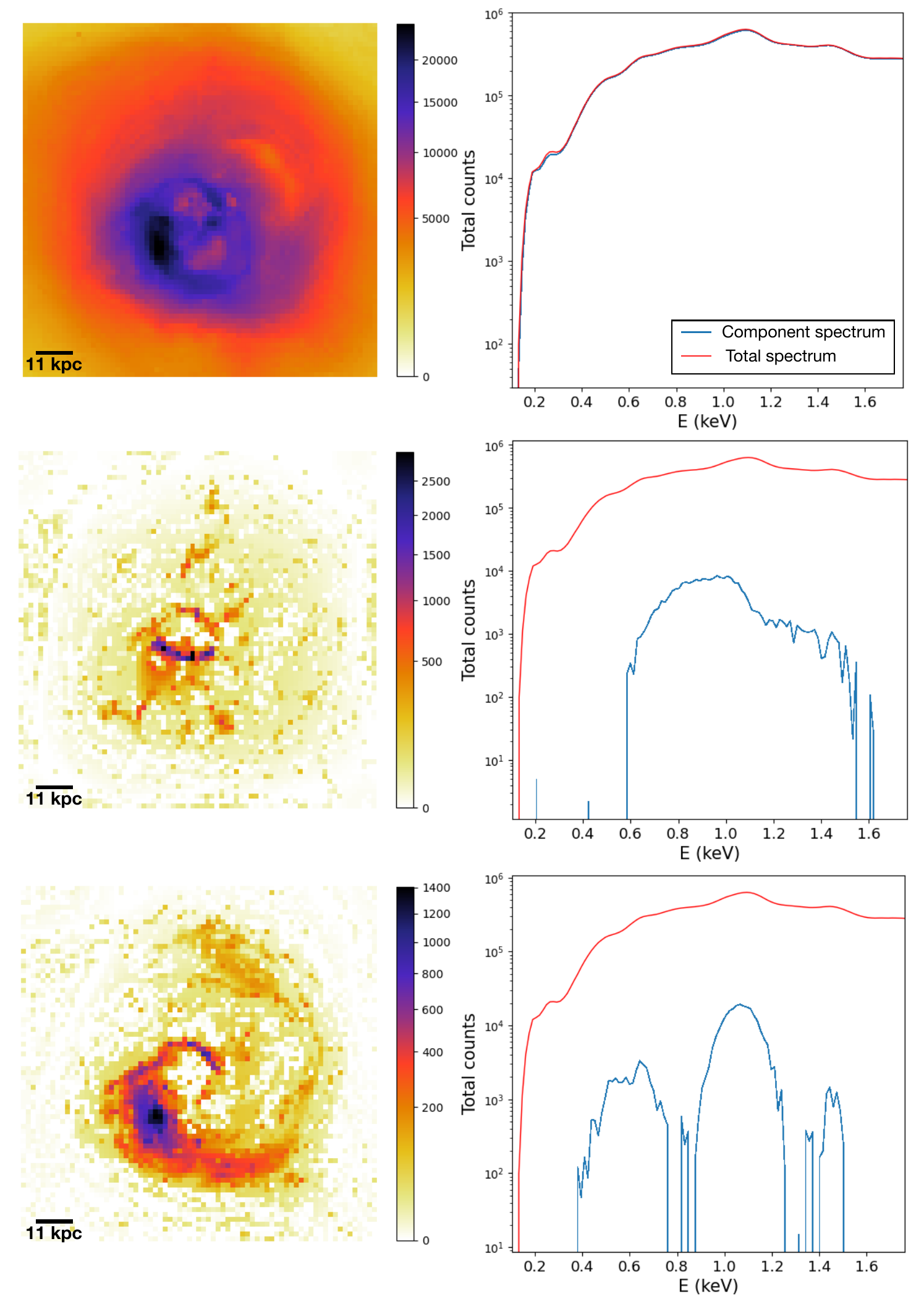}
\caption{Three components found by the pGMCA algorithm between $0.1$ and $1.8$ keV on the spectrally native, spatially rebinned data. On the left, spatial maps in square root and on the right, the associated integrated spectrum.}
\label{fig:pGMCA-results}
\end{figure}
\begin{figure*}[ht!]
\centering
\includegraphics[width=17cm]
{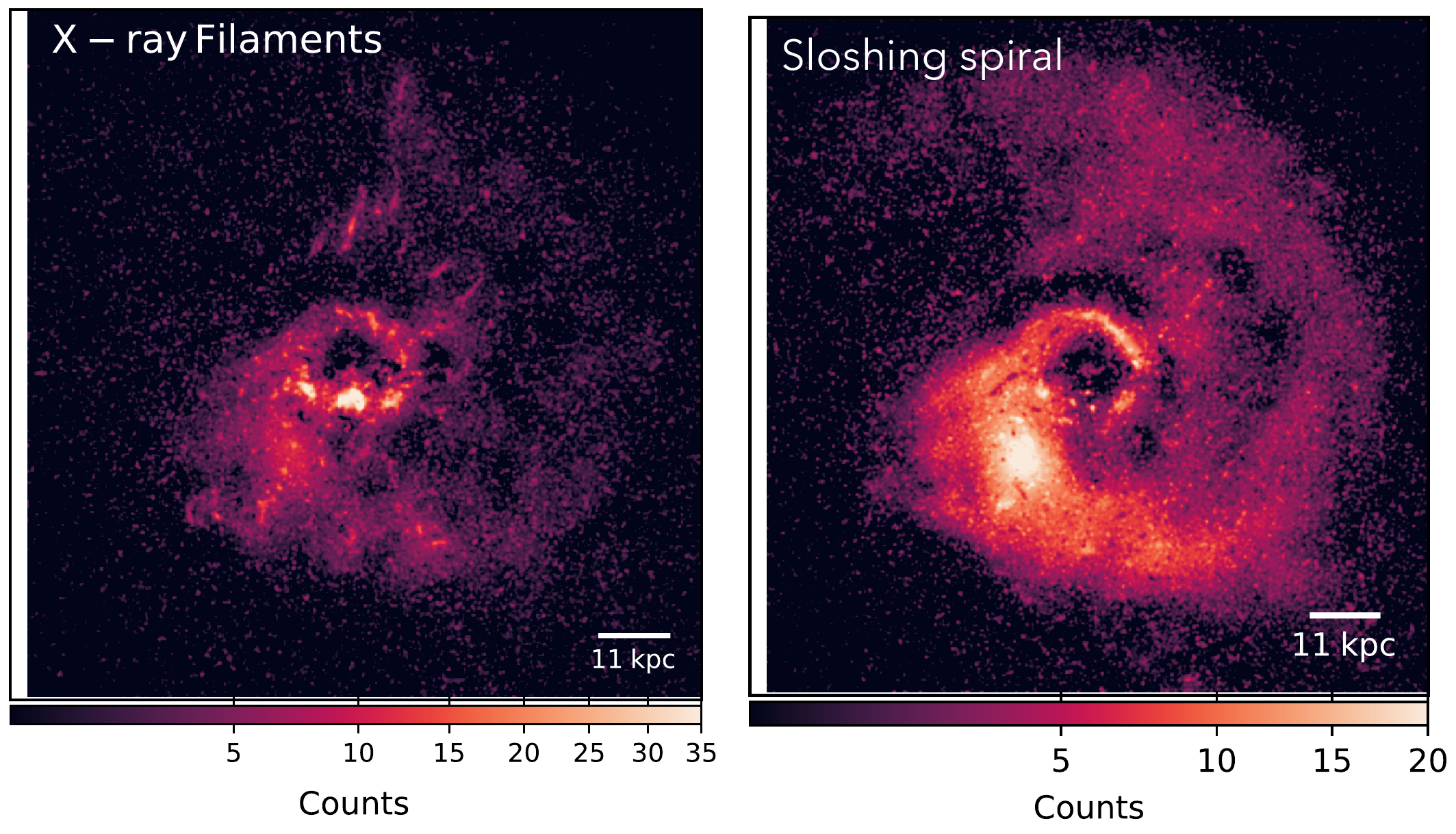}
\caption{Maps of the X-ray filaments and cold sloshing front. On the left, map of the X-ray filaments found by pGMCA between $0.5$ keV and $4.5$ keV on the spectrally rebinned, spatially native data, smoothed. On the right, map of the cold sloshing front retrieved by pGMCA between $0.5$ keV and $4.5$ keV on the spectrally rebinned, spatially native data, cleaned, smoothed For both images, the scale is in square root.}
\label{fig:both-images}
\end{figure*}

The application of pGMCA on the spatially rebinned X-ray data on a $0.1$ - $1.8$ keV energy range yielded three meaningful components, shown in Fig. \ref{fig:pGMCA-results}. The main component, which represents 98\% of the total counts, closely resembles the unprocessed \textit{Chandra} image. As previously stated, the other two components are faint but spatially and morphologically distinct and correspond to the X-ray counterparts to the H$\alpha$ filaments, and to the sloshing spiral. As pGMCA does not use any physical of instrumental information, the reconstructed spectra are not always physically meaningful; in particular, the spectra associated with faint components are sometimes poorly reconstructed. It is the case here: the spectra associated with the filaments and the sloshing spiral are not meaningful as such, but they indicate that these morphologically distinct components may also be spectrally distinct, corresponding to low-energy emission, with spectra peaking below $1.8$ keV. We also applied the pGMCA algorithm to the spectrally rebinned but spatially-native data in the $0.5$ - $4.5$ keV energy range, to obtain better-resolved images. The different components obtained from the pGMCA method are shown in Fig. \ref{fig:both-images}.

\subsection{Spectral analysis - a template fitting method in X-rays}

Here we introduce a template three-dimensional (RA, Dec, energy) fitting method for X-rays, inspired by gamma-ray analysis such as the one used for Fermi-LAT or Cherenkov telescopes \citep[i.e. using tools such as \textit{gammapy}\footnote{ \url{https://docs.gammapy.org/}} that is used in this study,][]{Donath_2023}. This method performs a global fit of the spatial and spectral dimensions of the data which requires to define both a spectral and a spatial model for each relevant physical component. While at gamma-rays energies most of the sources can be modeled by simple geometrical models (point source, Gaussians, disks, etc), this is not applicable in X-rays due to the drastically better PSF with respect to gamma-rays. Therefore the main challenge to perform a three-dimensional analysis in X-rays is to define meaningful spatial model. 

We used the information extracted from our imaging analysis above as a spatial constraint for our spectral analysis. Our approach to tackle this challenge consists of using the spatial information derived from our component separation algorithm in previous sections and shown in Fig. \ref{fig:pGMCA-results}, but to replace the poorly reconstructed spectra with physically meaningful models. Note that one limitation of this method is that it assumes that the spectral properties do not vary spatially within a given spatial template. This limitation is partially lifted later in this section by performing a fit in different annular rings.
We therefore defined our spatial models as the sum of three absorbed \texttt{apec} models from \textit{Xspec} \citep[][]{Xspec}, each multiplied by the spatial template of the corresponding component extracted by pGMCA. We tied the metallicities of each \texttt{apec} model, and undertook the fit in the $0.5$ - $4.5$ keV energy range. 
% The main idea is to fit the whole data cube with a sum of three absorbed \texttt{apec} models from \textit{Xspec} \citep[][]{Xspec} to model the fully ionized plasma, each multiplied by one of the components extracted by pGMCA as a spatial template. We tied the metallicities of each \texttt{apec} model, and performed the fit on the $0.5$ - $4.5$ keV energy range.

For this analysis, we kept the $14.6$ eV native energy binning from the merged \textit{Chandra} observations, but we rebinned the cube and the spatial templates $5$ times spatially to a $2.5$" spatial bin size in order to increase the statistics in each spatial bin. We cleaned the sloshing spiral and filament images used as spatial templates, denoising them by removing the coarsest wavelet scale. For the filament template, we applied a spatial mask defined using the contours of the H$\alpha$ image from \cite{Conselice_2001} (shown in Fig. \ref{fig:filaments-boxes-profiles}, top right) to avoid residual contamination.
% and we applied a mask defined using the contours of the H$\alpha$ image from \cite{Conselice_2001} (shown in Fig. \ref{fig:filaments-boxes-profiles}, top right) on the filaments image to avoid residual contamination. 
The use of a mask defined on the H$\alpha$ filaments is possible in this case because these structures are co-spatial, as will be discussed in Sec. \ref{sect:filaments}.

The instrumental background was accounted for using the \texttt{blanksky} routine from \textit{CIAO 4.16.0}, and the column density was fixed to $1.4 \times 10^{21}$ cm$^{-2}$, using the value from the LAB survey \citep[][]{Kalberla2005}. As the templates are extracted from the data and not from simulations, they are already impacted by the instrument's Point Spread Function (PSF) in the same way as the data cube we are attempting to fit, so there is no need to add anything to take the PSF into account. The model was defined using the \texttt{SkyModel} function from \textit{gammapy}. The combined RMF and ARF files given by \textit{Xspec}'s \texttt{specextract} are used in the fit. The fit was performed using the \texttt{simplex} minimization algorithm, and we estimated errors using \textit{Ultranest} \citep[][]{buchner2021}. The spatial constraint being extremely strong with the template fitting method, the errors derived this ways are very small. To study the dependence of the fitted parameters on the distance from the center, we undertook the template fit in three separate annular regions shown in the top-left panel of Fig. \ref{fig:filaments-boxes-profiles}, and Fig. \ref{fig:cold-front}. The best fit parameters in the three regions are presented in Table \ref{table:3d}. We also fitted the three regions together, and the results are shown in the same Table, and in Fig.\ref{fig:fit-spectrum}. For the fits in the combined regions, the filament abundances were tied to that of the main component, but the sloshing spiral abundances were left free. For the fits in the three individual regions, we had to proceed differently because of the lower statistics:  a first fit was performed with all the abundances tied, then the sloshing spiral abundance was freed and the temperature frozen to the best value from the initial  fit.

\section{Discussion}
\label{sect:discussion}

\subsection{X-ray counterparts to the H$\alpha$ filaments}
\label{sect:filaments}

Multiphase filaments surrounding the central galaxy of cooling flow clusters can result from hot gas condensation triggered by AGN feedback \citep[e.g.][]{Edge_01,olivares19,Russell19}. In Perseus, the central galaxy NGC 1275 is surrounded by H$\alpha$ filaments extending over $100$kpc \citep[][]{Lynds70,Conselice_2001,Fabian_2008}. X-ray counterparts to these H$\alpha$ filamentary structures were observed in the outer regions of the cluster more than twenty years ago \citep[][]{Fabian03-filaments}, but until recently, no quantitative correlation had been established between the X-ray and H$\alpha$ surface brightness. In \cite{Olivares_2025}, the pGMCA algorithm allowed for the extraction of X-ray counterparts to the H$\alpha$ filaments from the \textit{Chandra} observations of seven clusters, including Perseus. This enabled the discovery of a tight correlation between the X-ray and H$\alpha$ surface brightness in these structures, providing a strong argument in favor of condensation of the hot ICM through Chaotic Cold Accretion models \citep[e.g.][]{Gaspari_2013}.

\begin{figure*}[ht!]
\centering
\includegraphics[width=18cm]
{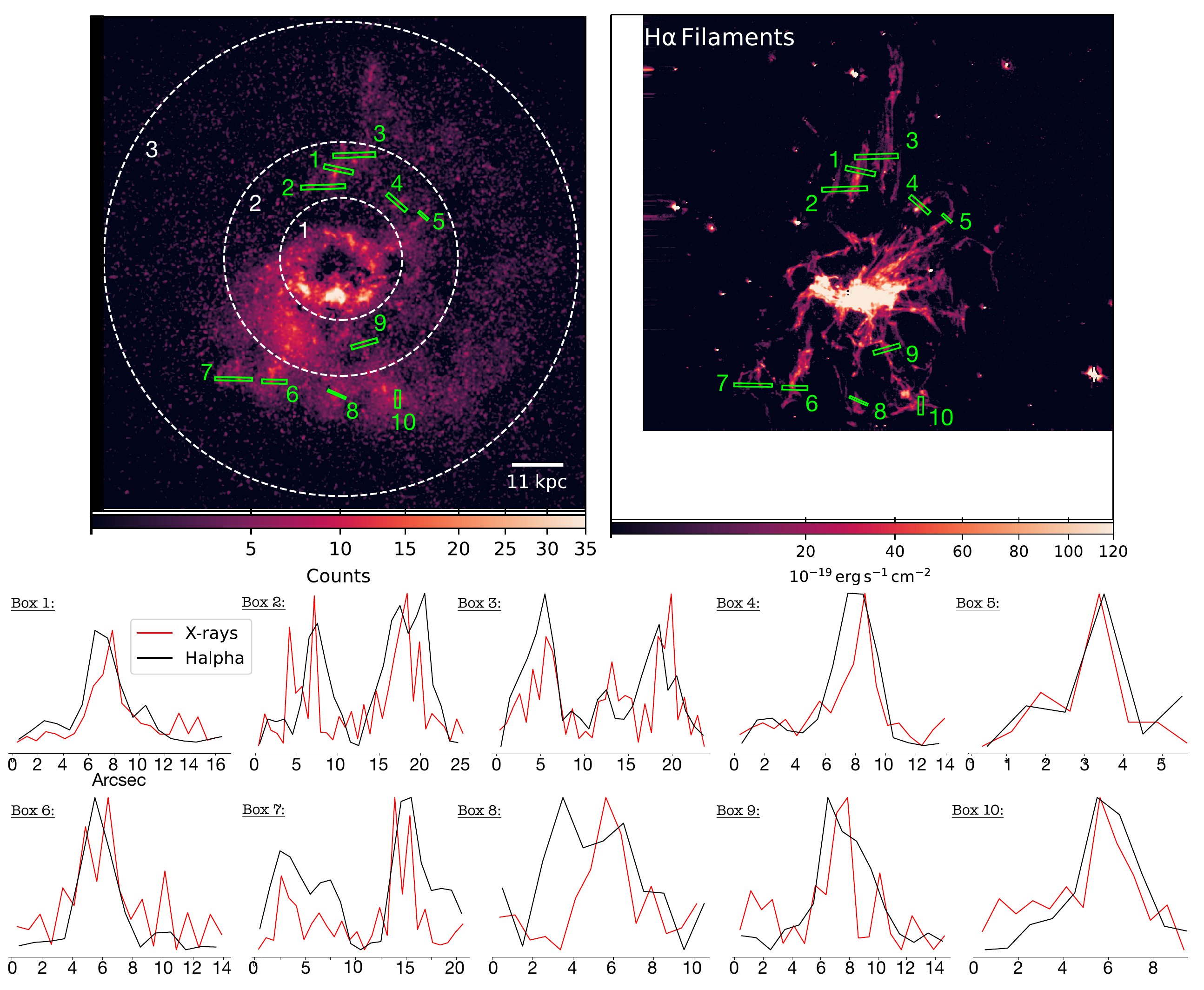}
\caption{On top, distribution of the X-ray and H$\alpha$ filaments. On the left, map of the X-ray filaments from Fig. \ref{fig:both-images}. On the right, H$\alpha$ filaments observed by the WIYN 3.5-m telescope, first presented in \cite{Conselice_2001}. Superimposed, the green box regions we defined to extract the filaments profiles. The white annular regions are the ones used in our template fit (see results in Table \ref{table:3d}). For both images, the scale is in square root. On the bottom, normalized profiles extracted from the X-ray filamentary structures found in the the X-ray and H$\alpha$ filaments in red and black, respectively. The X-ray filaments are obtained with the pGMCA method, while the H$\alpha$ come from the WIYN 3.5-m telescope.}
\label{fig:filaments-boxes-profiles}
\end{figure*}

In Fig. \ref{fig:both-images} (right panel), we present a map of the X-ray filaments with significantly improved resolution compared to that shown in \cite{Olivares_2025}, and in the top panels of Fig. \ref{fig:filaments-boxes-profiles}, we show this map together with the H$\alpha$ filaments from \cite{Conselice_2001}. In the bottom panel of Fig. \ref{fig:filaments-boxes-profiles}, we show a comparison of the X-ray and H$\alpha$ filaments' widths along ten different profiles: the regions corresponding to the profiles are illustrated in the top panels of Fig.~\ref{fig:filaments-boxes-profiles}. 

Our comparison shows that the X-ray and H$\alpha$ filaments are co-spatial and have apparent similar widths of about $700$ - $1400$ pc when observed by \textit{Chandra} and \textit{WIYN 3.5-m}, which have comparable spatial resolutions of $0.5$" and $0.7$", respectively. Unfortunately, these resolutions are not good enough to constrain further the actual widths of these filaments, that were observed to be around $70$ pc in H$\alpha$ by Hubble Space Telescope \citep[\textit{HST}][]{Fabian_2008}. In a future study, a more comprehensive assessment of the X-ray and H$\alpha$ co-spatiality could be carried out, and include \textit{HST} profiles.

%\begin{figure*}[ht!]
%\centering
%\includegraphics[width=17cm]
%{images/filaments-boxes-new.pdf}
%\caption{Distribution of the X-ray and H$\alpha$ filaments. On the left, map of the X-ray filaments from Fig. \ref{fig:both-images}. On the right, H$\alpha$ filaments observed by the WIYN 3.5-m telescope, first presented in \cite{Conselice_2001}. Superimposed, the green box regions we defined to extract the filaments profiles. The white annular regions are the ones used in our template fit (see results in Table \ref{table:3d}). For both images, the scale is in square root.}
%\label{fig:filaments-boxes}
%\end{figure*}

%\begin{figure*}[ht!]
%\centering
%\includegraphics[width=17cm]
%{images/profiles.pdf}
%\caption{Normalized profiles extracted from the X-ray filamentary structures found in the the X-ray and H$\alpha$ filaments in red and black, respectively, from the boxes presented in Fig. \ref{fig:filaments-boxes}. The X-ray filaments are obtained with the pGMCA method, while the H$\alpha$ come from the WIYN 3.5-m telescope.}
%\label{fig:filaments-profiles}
%\end{figure*}

The best fits obtained with our template fitting method shown in Table \ref{table:3d} indicate temperatures between $0.70$ keV and $0.80$ keV for the filaments. These temperatures are similar to that of the lower components from the \cite{Fabian11} and \cite{walker15} studies, where X-ray spectra extracted from the filaments were fitted with two absorbed \texttt{apec} models: the twelve samples from \cite{walker15} give an average lower temperature of $0.79$ keV. Interestingly, in our fits, the filaments appear to be somewhat warmer in the inner regions. Despite the difference in methodology and the wide fitting regions, the temperatures we retrieve for the X-ray filamentary structures appear consistent with past studies.

\begin{table*}
    \centering
     
    \renewcommand{\arraystretch}{1.4}
    \begin{tabular}{c c c c c c}

Component & Parameter & All regions & Region 1 & Region 2 & Region 3 
 \\ 
\hline
Main & kT (keV) & 3.18   & 3.30    & 3.07  & 3.22  
\\ 
 & Abundance & 0.77  & 0.84    & 0.67  & 0.75  
\\ 
 & norm & 1.68  & 1.61   & 1.68  & 1.64  
\\ 
\hline
Filaments & kT (keV) & 0.76  & 0.80   & 0.78  & 0.70  
\\ 
 & Abundance & 0.77 (tied) & 0.84 (tied)  & 0.67 (tied) & 0.75 (tied) 
\\ 
 & norm & 0.001  & 0.002   & 0.0004  & 0.001  
\\ 
\hline
Sloshing spiral & kT (keV) & 1.37  & 1.41   & 1.37  & 1.29  
\\ 
 & Abundance & 2.1  & 2.98    & 1.65  & 0.71  
\\ 
 & norm & 0.006  & 0.007    & 0.008  & 0.019  
\\ 
\hline 
\end{tabular}
\caption{\label{tab:data-description}Best fits obtained with our template fitting method to describe the three components found by pGMCA. Each component is described with an absorbed \texttt{apec} model multiplied by one of the three maps extracted by the algorithm as a spatial template. The nH value is fixed at $1.4 \times 10^{21}$ cm$^{-2}$, and the redshift $z$ at $0.018$. The filaments abundances are tied with that of the main component. The fit was performed in three separate annular regions, shown in Fig. \ref{fig:filaments-boxes-profiles}, left, and Fig. \ref{fig:cold-front}, and in all the regions combined. The errors derived using the \textit{Ultranest} method are of the order of $0.02$ keV for the temperatures, and $0.1$ for the norms. The abundances being fitted separately, the errors we could retrieve were not meaningful.}
\label{table:3d}
\end{table*}

\subsection{Sloshing spiral}
\label{sect:coldfront}

Cold fronts were first observed by \cite{Markevitch_1999} as sharp edges in X-ray surface brightness similar to shock fronts, but at pressure equilibrium. A proposed mechanism for their formation is that of gas ``sloshing'' in the gravitational potential well of a cool-core cluster, typically initiated by the passage of a subcluster or a galaxy. The cold gas from the center is being pushed away from the potential minimum, causing gas motions that evolve over time into large-scale spiral-shaped structures \citep[e.g. simulations from][]{Ascasibar_2006,ZuHone_Roediger_2016,Bellomi_2024}. 

\textit{Chandra} observations have led to the discovery of cold fronts in a number of galaxy clusters \citep[][]{Markevitch2000,Vikhlinin2001}. In Perseus, observations have revealed both an inner cold front and an outer, ancient one at distance of a 730\,kpc from the center \citep[][]{Walker_2018}. The recent weak lensing detection of a low-mass subcluster halo to the West of the main Perseus core provides a plausible generation mechanism for the spiral structure and associated cold fronts \citep{hyeonghan}.

Similar to the X-ray filaments, direct imaging of the inner cold front and spiral structure is made difficult by the hot ICM emission from the surrounding medium. In the left-hand panel of Fig. \ref{fig:both-images}, we present the first detailed, unpolluted map of the inner cold front and spiral structure in Perseus, obtained with the pGMCA algorithm. It strongly resembles the temperature maps from \cite{Sanders05} (Fig. 19) and \cite{Fabian_2006} (Fig. 4). The component extracted by the pGMCA algorithm is associated with a low energy, high-metallicity spectrum, which is confirmed by our template fits, giving temperatures around $1.3$ - $1.4$ keV and higher abundances than in either the filaments or the surrounding medium in all three regions. This component corresponds to the spiral feature detected between $2$ and $4$ keV in previous temperature maps \citep[e.g.][]{Sanders05,Fabian_2006,Sanders07,Walker_2018}.
% However, the component extracted by the pGMCA algorithm is associated with a surprisingly low energy spectrum, which is confirmed by our template fits, giving temperatures around $1.3$ - $1.4$ keV in our three regions. Past studies did not observe the cold front directly, but it appeared as a spiral feature mainly between $2$ and $4$ keV in temperature maps \citep[e.g.][]{Sanders05,Fabian_2006,Sanders07,Walker_2018}.
The template fitting method is sensitive to initial conditions, so we also applied a more traditional approach. We extracted a spectrum from a region devoid of filaments and where the sloshing spiral is relatively bright, and fitted it with \textit{Xspec}. The results are presented in Appendix \ref{ap:xspec}. The temperatures and abundances obtained this way are consistent with our findings with the template fitting method. 
% , and confirm that the low energy emission from the cold front is extremely faint and could easily be overlooked with a classical approach.

% The pGMCA being efficient in disentangling morphologically and spectrally distinct components, it was able to capture an extremely faint part of the cold front emitting below $1.5$ keV that standard analysis would not detect. Similarly, the spatial information used in our template fitting method allowed for the modeling of this elusive component and determine a temperature of $\sim 1.35$ keV on average. To our knowledge, this is the first direct observation of a lower energy emission from the cold front in X-rays (MAKE SURE OF THIS).

The combination of higher abundances, lower temperatures, and distinct morphology found in the sloshing spiral may explain its robust retrieval in the blind source separation algorithm. The observed higher abundances in the sloshing spiral may be due to movement of the BCG in the center of the potential as a result of the recent merger. In turn, the spiral produces an offset in the peak of the abundance relative to the center of the X-ray emission. This offset, as well as AGN feedback, may contribute to the observed phenomenon of the central abundance dip seen in Perseus and other systems \citep{Sanders07, mernier17}.

%In Appendix \ref{ap:axis}, we evaluated the capabilities of the proposed Advanced X-ray Imaging Satellite \citep[\textit{AXIS},][]{2023SPIE12678E..1ER}, in comparison to those of \textit{Chandra} to study the observed features in  Perseus cluster core.

\section{Conclusions}

In this paper we report direct imaging and spectroscopic observations of the X-ray counterparts to the H$\alpha$ filaments and of the sloshing spiral and associated cold fronts in Perseus, obtained using a blind source separation method applied to \textit{Chandra} observations. The X-ray counterparts to the H$\alpha$ filaments are co-spatial with the filaments, and their projected widths are in remarkably good agreement. We introduced a template fitting method for X-rays using the extracted maps as spatial templates, which allowed us to retrieve temperatures and abundances for the filaments and the sloshing spiral in three annular regions. The filament temperatures obtained using this method are in good agreement with previous analyses. The sloshing spiral exhibits low temperatures and higher abundances than either the filaments or the surrounding medium. This component may correspond to highly enriched gas that is entrained by the movement of the BCG due to the ongoing merger event. These results illustrate the potential of blind-source separation methods when applied to deep high-angular resolution X-ray data. In the future, the Advanced X-ray Imaging Satellite \citep[\textit{AXIS},][]{2023SPIE12678E..1ER}, endowed with a substantially larger effective area, will yield a photon statistics higher by a factor of $\sim$20. Its more stable PSF across the field of view, combined with the power of the pGMCA and template fitting methods presented here, could reveal larger-scale structures further away from the cluster's core emission.

%While the $\sim 0.7-0.9$~keV temperatures associated with the filaments are consistent with past studies, the $\sim 1.3$ keV temperature we find for the cold front is significantly lower than the $\sim 2-4$~keV that temperature maps of Perseus usually associate with this structure. To our knowledge (WE SHOULD MAKE SURE OF THAT), our results thus constitute the first direct observation of a $\sim 1.35$ keV low energy emission from the cold front.

\begin{acknowledgments}
This work made use of Gammapy \citep{Donath_2023}, a community-developed Python package.
VO acknowledges support from the DYCIT ESO-Chile Comite Mixto PS 1757, Fondecyt Regular 1251702.\\

This paper employs a list of Chandra datasets, obtained by the Chandra X-ray Observatory, contained in~\dataset[DOI: CDC478]{https://doi.org/10.25574/cdc.478}.
\end{acknowledgments}

\appendix

\section{Template fitting limitations and \textit{Xspec} Analysis of the cold front}
\label{ap:xspec}

\begin{figure}%[ht!]
\centering
\includegraphics[width=8cm]
{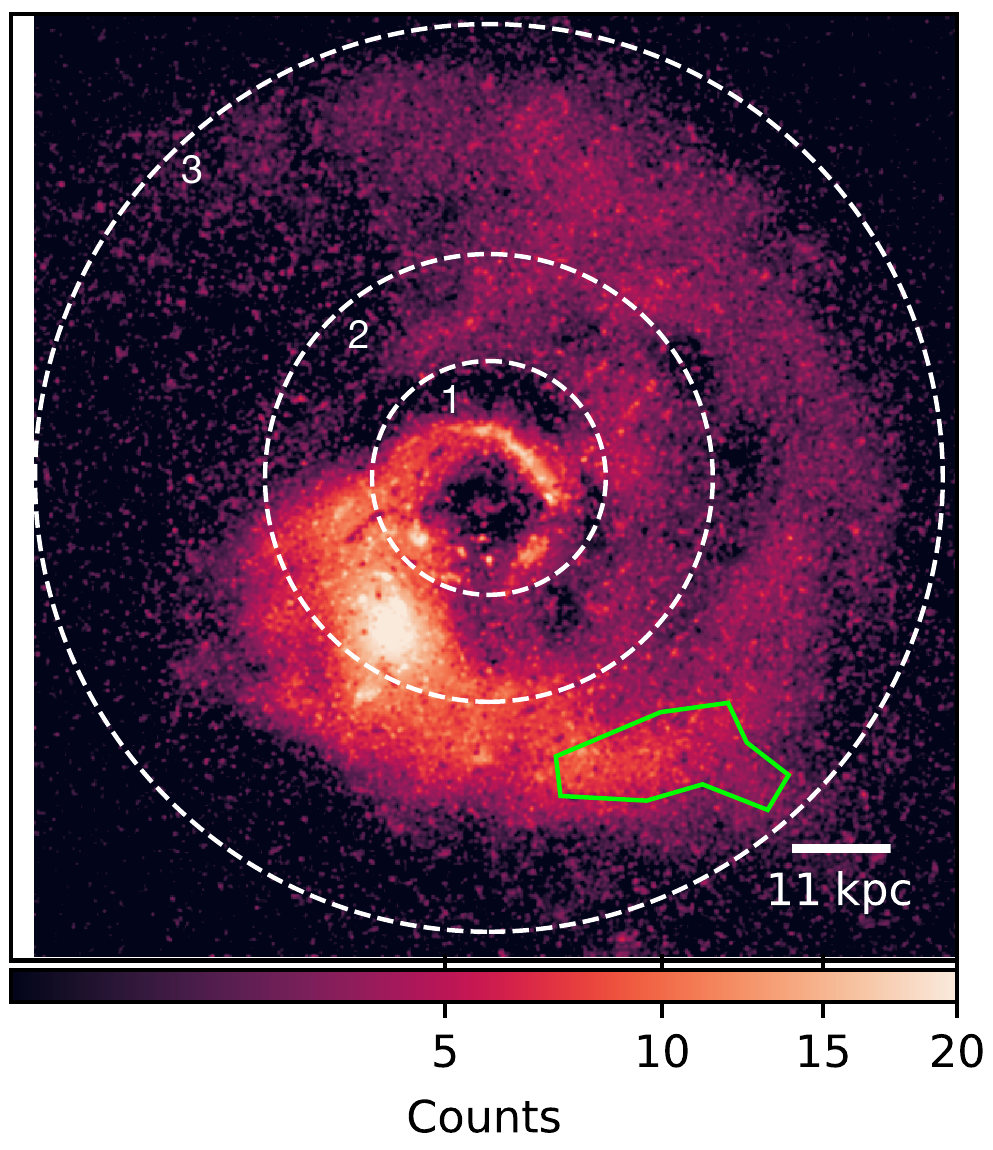}
\caption{Image of the sloshing spiral from Fig. \ref{fig:both-images} superimposed with spectral analysis regions. The white annular regions are the ones used in our template fit (see results in Table \ref{table:3d}), the green region is the one used in the traditional \textit{Xspec} analysis (see Table \ref{tab:xspec}).}
\label{fig:cold-front}
\end{figure}

\begin{figure*}[ht!]
\centering
\includegraphics[width=17cm]
{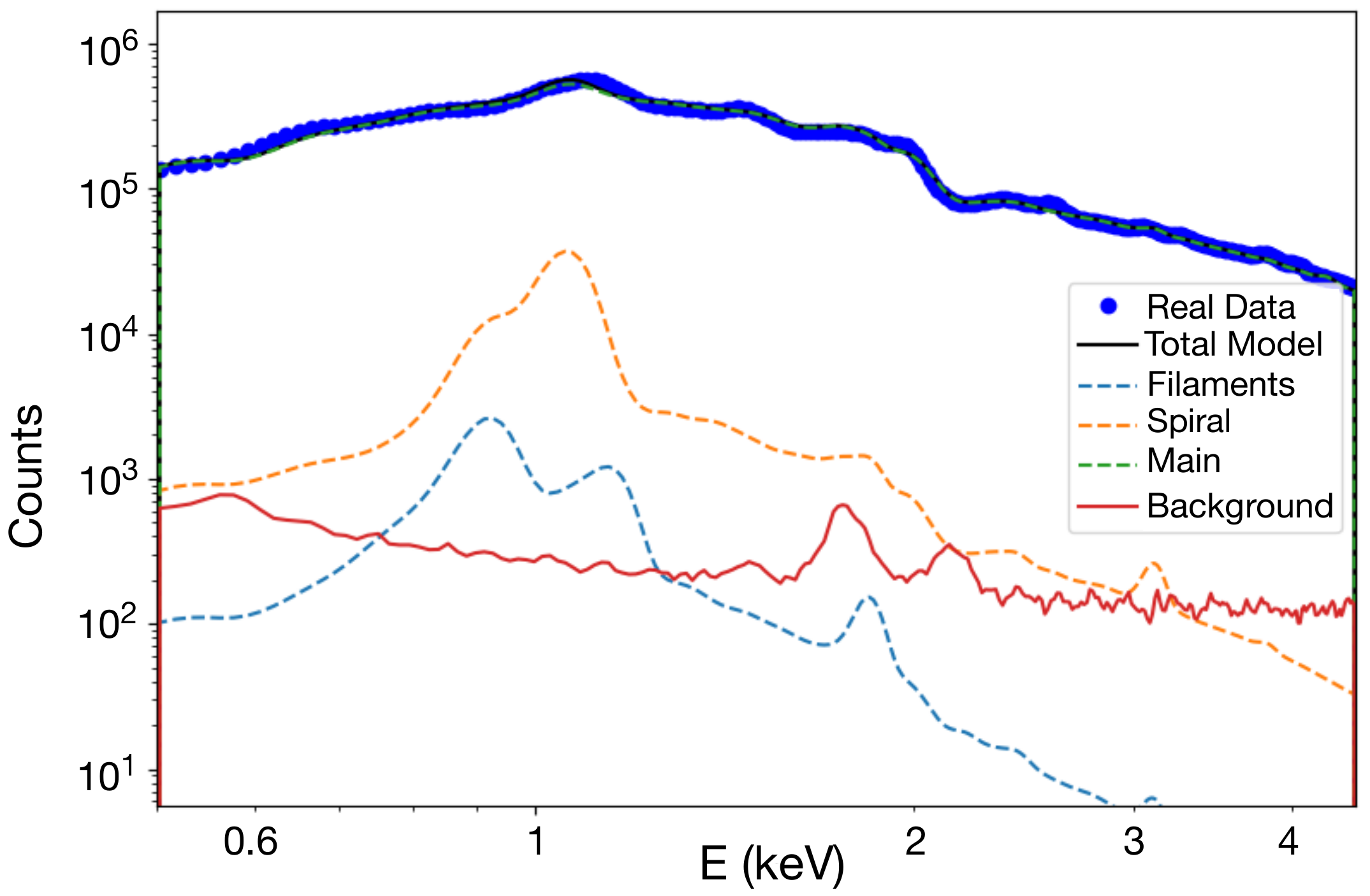}
\caption{Best fit for the template fitting method on all annular regions combined. The regions are shown in  Fig. \ref{fig:filaments-boxes-profiles}, left, and Fig. \ref{fig:cold-front}.}
\label{fig:fit-spectrum}
\end{figure*}

The template fitting method we developed for X-ray astronomy allows for the study of faint components by using as much information as possible from the three-dimensional $(x,y,E)$ data given by spectro-imagers. However, using a single spectral model to fit all the pixels from a specific physical component is a strong assumption, and the results of the fit are sensitive to initial conditions. In Fig. \ref{fig:fit-spectrum}, we can nonetheless see that the best fit we obtain with this method yields morphologically distinct spectra associated with each template.

To challenge these results, we also applied a more traditional approach, by extracting a spectrum from the green region shown in Fig. \ref{fig:cold-front} and fitting it with a multi-component spectral model in \textit{Xspec}. This region was selected for the relative brightness of the sloshing spiral and because it is devoid of filaments, in order to avoid pollution. We first fitted the extracted spectrum between $0.5$ and $4.5$ keV  with a single absorbed \texttt{apec} model, then with two absorbed \texttt{apec} models with tied or free abundances. The results of our fits are presented in Table \ref{fig:cold-front}, and show that the extracted spectrum is significantly better described by two absorbed \texttt{apec} models, one associated with a temperature of $\sim 3.15$ keV, and one with a temperature of $\sim 1.55 \pm 0.2$ keV. This is consistent our findings using the template fitting method. Freeing the abundances of the two \texttt{apec} models does not improve the fit, and the abundance of the second \texttt{apec} is poorly constrained. This is consistent with with the template fitting results from Region 3, where the green extracting region is located.

We tried the same method on a larger $0.5$ - $7$ keV energy range, but the best fit with two absorbed \texttt{apec} models with linked or free abundances gave temperatures of $\sim 2.6$ keV and of $\sim 5$ keV. The $\sim 5$ keV temperature is consistent with previous studies for the sloshing spiral, but is significantly higher than the $1.3$ keV found by our template fitting method. It shows that the low energy emission from the sloshing spiral is extremely faint and could easily be overlooked with a classical approach.

\begin{table}
    \centering

    \renewcommand{\arraystretch}{1.4}
    \begin{tabular}{c c c}
    
Model & Parameter & Best fit 
 \\ 
\hline
\texttt{phabs} & nH ($10^{22}$ cm$^{-2}$) & 0.13 $\pm$  1.0e-3 
\\ 
\texttt{*} & kT (keV) & 3.03 $\pm$  1.7e-2 
\\ 
\texttt{apec} & Abundance & 0.54 $\pm$  9.7e-3 
\\ 
 & norm & 5.48e-3 $\pm$  2.9 e-5
\\ 
\cline{2-3}
 & $\chi^2$ & 602.68 (269 d.o.f)
 \\ 
\hline
\texttt{phabs} & nH ($10^{22}$ cm$^{-2}$) & 0.13 $\pm$  1.0e-3 
\\ 
\texttt{*}  & kT$_1$ (keV) & 3.17 $\pm$  5.1e-2 
\\ 
\texttt{(apec1+apec2)} & Abundance$_1$ & 0.55 $\pm$  1.4e-2
\\ 
(linked abundances) & norm$_1$ & 5.25e-3 $\pm$  1.2e-4 
\\ 
 & kT$_2$ (keV) & 1.53 $\pm$  0.22 
\\ 
 & Abundance$_2$ & 0.55 (linked)
\\ 
 & norm$_2$ & 1.70e-4 $\pm$  1.4e-4 
\\ 
\cline{2-3}
 & $\chi^2$ & 548.94 (267 d.o.f)
 \\ 
 \hline
\texttt{phabs} & nH ($10^{22}$ cm$^{-2}$) & 0.13 $\pm$  1.7e-3 
\\ 
\texttt{*}  & kT$_1$ (keV) & 3.14 $\pm$  0.12 
\\ 
\texttt{(apec1+apec2)} & Abundance$_1$ & 0.52 $\pm$  4.2e-2
\\ 
(free abundances) & norm$_1$ & 5.37e-3 $\pm$  3.6e-4 
\\ 
 & kT$_2$ (keV) & 1.59 $\pm$  0.24 
\\ 
 & Abundance$_2$ & 3.53  $\pm$  42.2
\\ 
 & norm$_2$ & 3.26e-5 $\pm$  4.3e-4 
\\ 
\cline{2-3}
 & $\chi^2$ & 545.90 (266 d.o.f)
 \\ 
 \hline

\end{tabular}
\caption{\label{tab:data-description}Best fits obtained in {\it Xspec} to describe the green region of Fig. \ref{fig:cold-front}, with one or two absorbed \texttt{apec} models, with linked or free abundances. The redshift, $z$, is fixed at $0.018$. The fit was done between $0.5$ and $4.5$ keV.}
\label{tab:xspec}
\end{table}

\bibliography{perseus-bib}{}
\bibliographystyle{aasjournalv7}



\end{document}